\documentclass{appolb}
\usepackage{graphicx}
\usepackage{color}
\usepackage{soul}
\usepackage{hyperref}

\newcommand{\snn}[1]{$\sqrt{s_{_{NN}}}$ = #1\,GeV}

\begin{document}

\eqsec  

\title{Low $p_T$ direct photon production at RHIC measured with PHENIX 
\thanks{XXIXth International Conference on Ultra-relativistic Nucleus-Nucleus Collisions}
}
\author{Roli Esha (for the PHENIX Collaboration, \\
\href{https://doi.org/10.5281/zenodo.7430208}{https://doi.org/10.5281/zenodo.7430208})
\address{Department of Physics and Astronomy, Stony Brook University, \\
Stony Brook, NY 11790 USA.
}
}
\maketitle
\begin{abstract}
PHENIX has used the versatility of RHIC to map out low $p_{T}$ direct photon production as function of collision system size and beam energy. For systems with a size corresponding to a $dN_{ch}/d\eta$ larger than 20-30, we observe a large yield of direct photons, a large azimuthal anisotropy with respect to the reaction plane, and a characteristic centrality dependence of $dN_{\gamma}/dy \propto (dN_{ch}/d\eta)^{\alpha}$, with $\alpha \sim$ 1.2. \\
In this proceeding, we will present new results from Au+Au collisions at $\sqrt{s_{NN}}$ = 39, 62.4 and 200 GeV. After subtracting the prompt photon component, the inverse slope for the $p_{T}$ range from 1-2 GeV/$c$ is 250 MeV/$c$, but increases to about 400 MeV/$c$ for the range from 2 to 4 GeV/$c$. Within the experimental uncertainty, there is no indication of a system size dependence of the inverse slope. Furthermore, the system size dependence of the yield, expressed through the power $\alpha$, remains independent of $p_{T}$ over the entire observed range from 1 to 6 GeV/$c$. Like the large yield and azimuthal anisotropy, these features, while qualitatively consistent with the emission of thermal photons from the quark gluon plasma, elude a quantitative description through theoretical model calculations.
\end{abstract}
  
\section{Introduction}
Photons are a unique probe in understanding the properties of the hot and dense medium produced in relativistic heavy ion collisions as they do not interact with the medium strongly and carry unmodified information about the space-time evolution of the system. Direct photons are defined as those which do not come from hadronic decays. The direct photon spectrum is sensitive to the temperature of the medium and its measurement will help constrain initial conditions, sources of photon production, emission rates and the space-time evolution of relativistic heavy ion collisions.

All the photon sources can be classified into two categories -- decay photons, which constitute around 80-90\% of all the photons produced in heavy-ion collisions, and direct photons. The direct photons can further be split into two categories -- prompt and non-prompt. Prompt photons are the ones that come from sources similar to that in $p+p$ collisions and scale as the number of binary collisions. 
Beyond the known thermal sources from the Hadron Gas and the QGP phase, other examples of sources contributing to non-prompt direct photons are those from jet-medium interactions and from the pre-equilibrium state. Throughout the evolution of the system, it expands and cools. Hence, earlier phases are characterized by higher temperatures and likely to dominate the emissions at higher $p_{T}$.

The wealth of data and an optimized detector configuration has enabled PHENIX to measure direct photons over a large $p_T$ range, across 7 systems, 3 collision energies using 3 different methods, namely, the calorimeter method, the virtual photon method and the external conversion method. The availability of multiple data sets 
and different detector configurations
for the same beam energies provides for consistency checks.

This proceeding will focus on the external conversion method used 
for analyzing the direct photons from the years 2010~\cite{PHENIX:2022qfp} and 2014~\cite{PHENIX:2022rsx}.

\section{Direct photons}

\begin{figure}
\centerline{%
\includegraphics[height=5.25cm]{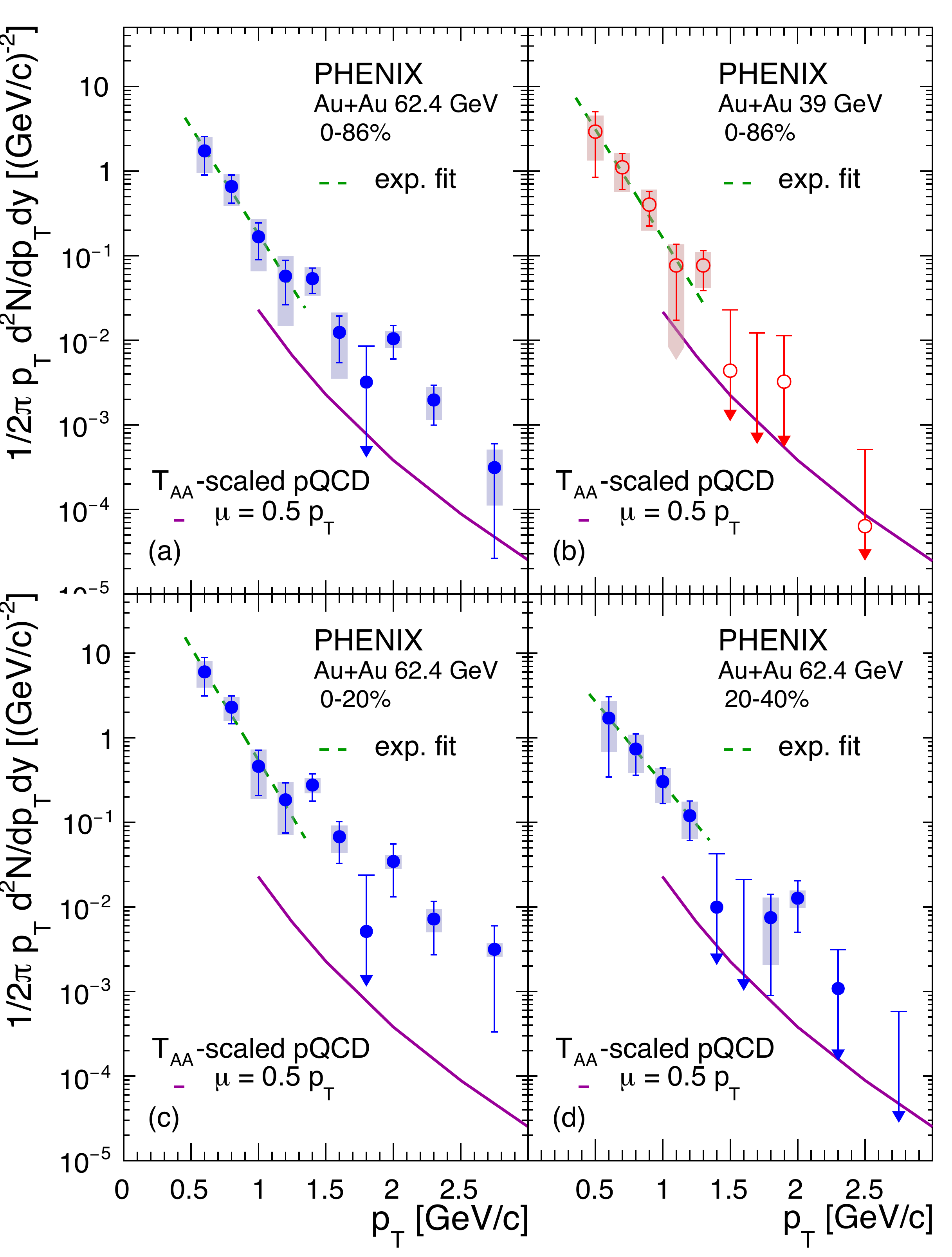}
\includegraphics[height=5.2cm]{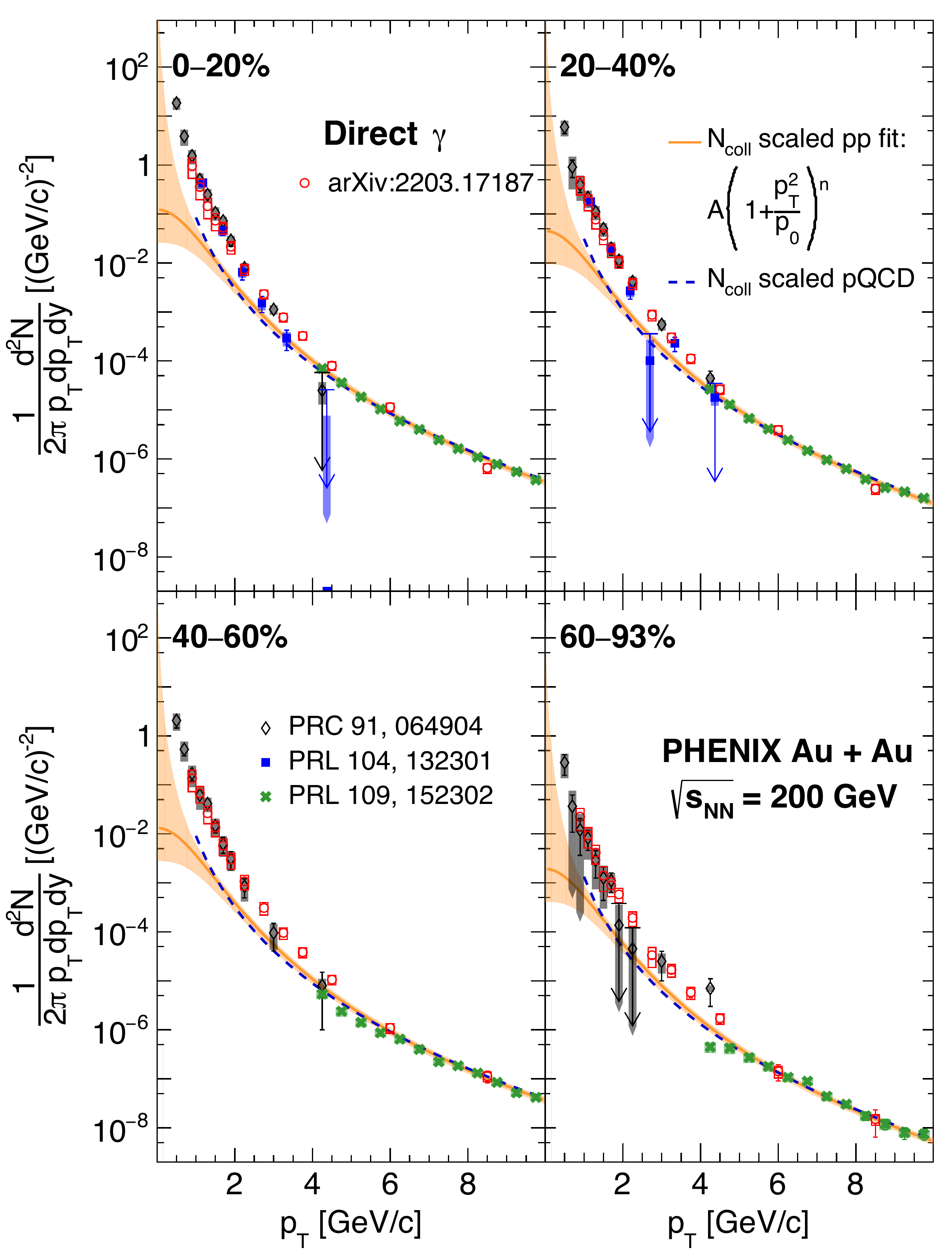}}
\caption{Invariant yield of direct photons as a function of $p_{T}$ for Au+Au at \snn{39} and \snn{64} (left) and for \snn{200} (right) for different collision centralities.}
\label{Fig:dp_spec}
\end{figure}

Photon conversions on the backplane of the Hadron Blind detector (HBD) were analyzed for Au+Au collisions 
recorded in 2010 at \snn{39} and \snn{62.4} and the corresponding direct photon spectra as a function of $p_{T}$ are shown in Fig.~\ref{Fig:dp_spec} (left) with the $T_{AA}$-scaled pQCD curve shown in solid line.

Significantly larger statistics of Au+Au collisions at \snn{200} were recorded in the year 2014. This allowed for a more differential measurement of direct photons. Instead of the HBD, which was removed, conversions in the layers of a new Silicon Vertex tracker, with a material budget of around 13\%, were analyzed. The spectrum for every 20\% collision centrality is shown in Fig.~\ref{Fig:dp_spec} (right). A good agreement with all the previous PHENIX measurements is observed. With the direct photon spectra established, the next step would be to understand the centrality dependence and the shape of the spectrum.

\begin{figure}
\centerline{%
\includegraphics[height=3.8cm]{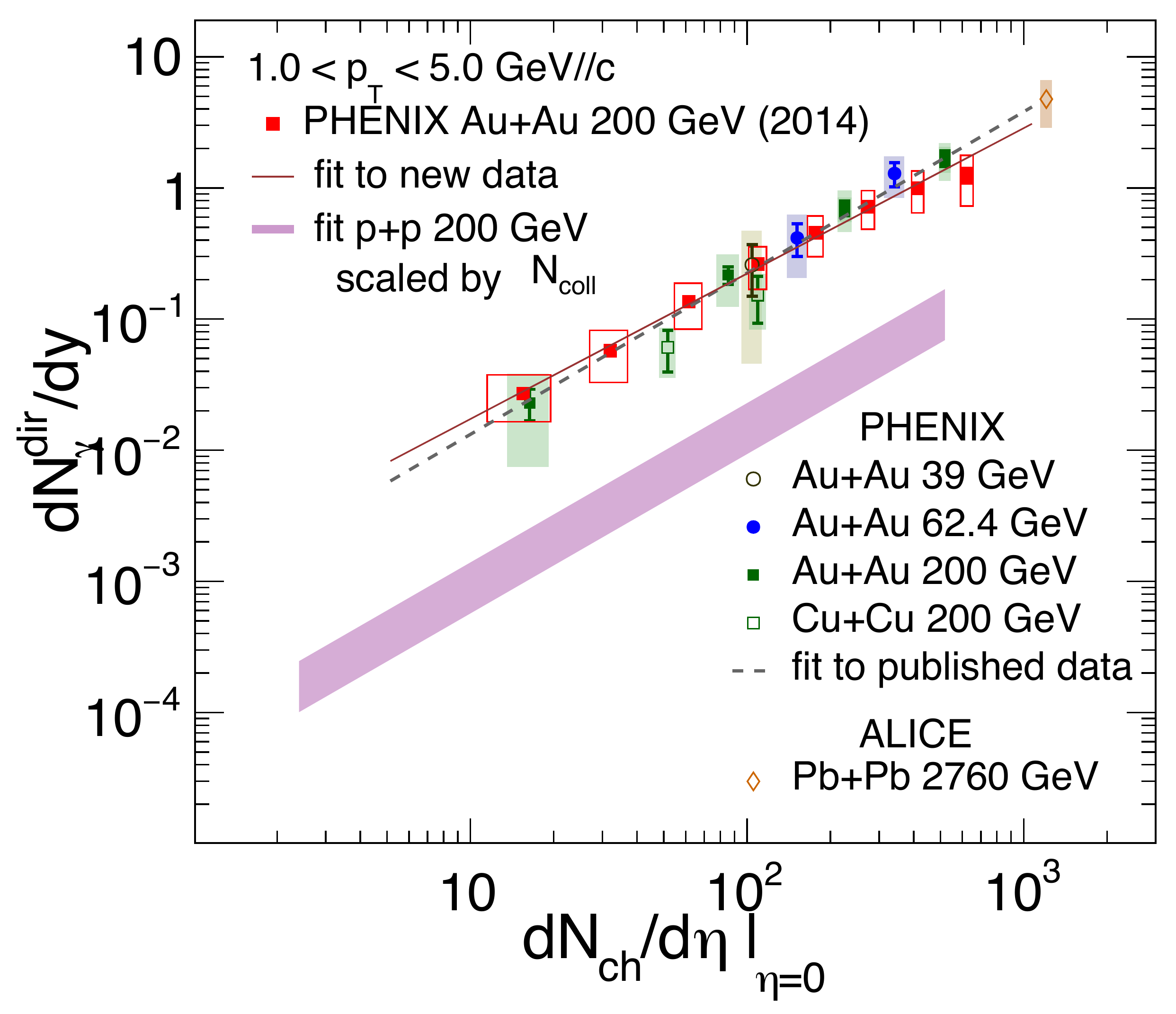}
\includegraphics[height=3.8cm]{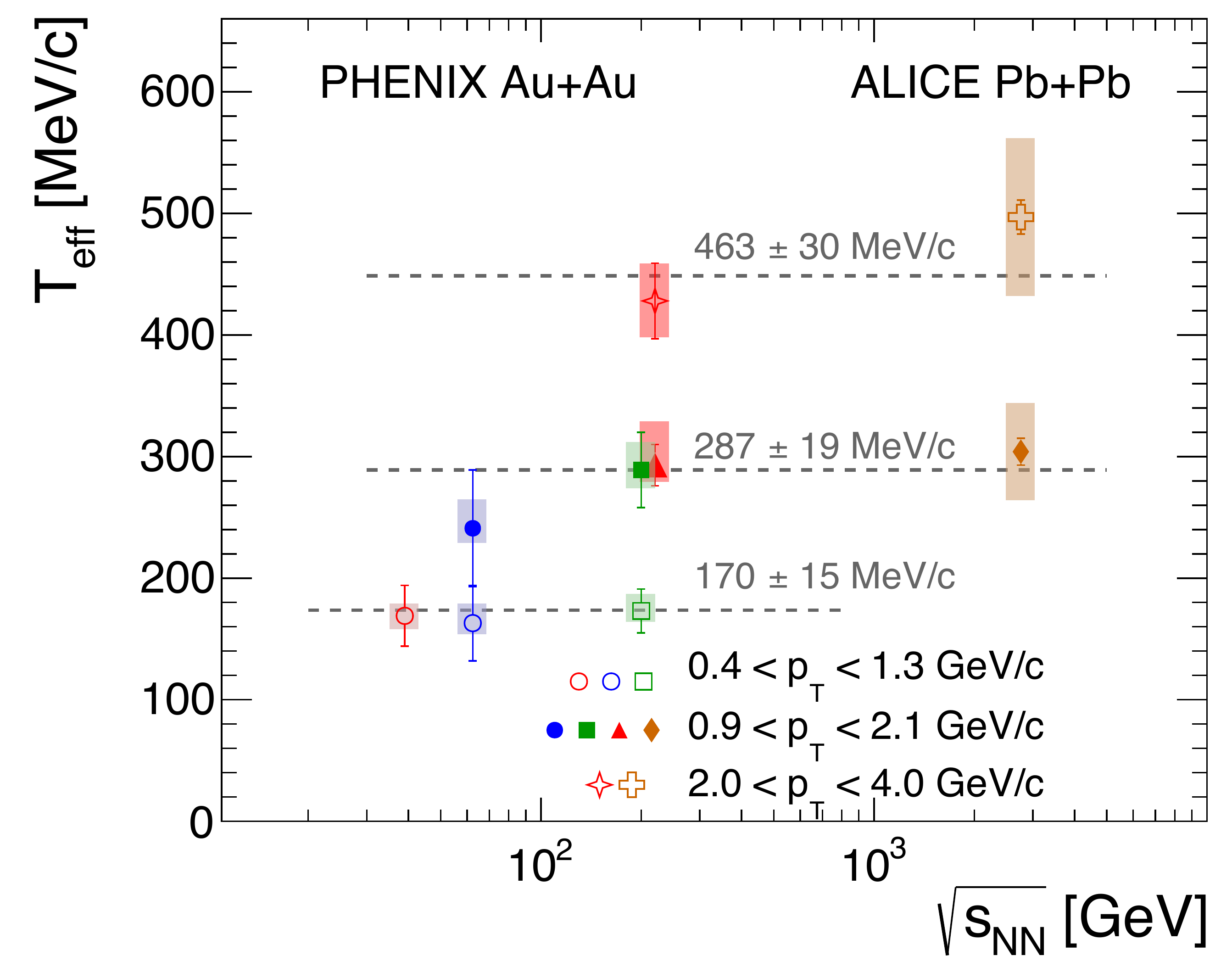}}
\caption{Integrated yield of direct photons as a function of system size (left) and the inverse slope of spectrum as a function of collision energy for different $p_{T}$ ranges (right).}
\label{Fig:dp}
\end{figure}

\textbf{Universal scaling}: In order to look deeper into centrality dependence, the integrated yields are analyzed in Fig.~\ref{Fig:dp} (left), where, the $p_{T}$-integrated direct photon yield is shown for various collision systems and energy spanning almost 2 orders of magnitude as a function of charged particle multiplicity at midrapidity. A universal scaling behaviour is seen in all A+A systems with a trend similar to that of scaled p+p, but with 10 times larger yield.

\textbf{Effective temperature}: The effective temperature is estimated as the local inverse slope of the spectrum. In order to better understand the similarity of low-$p_{T}$ direct photon spectrum across collision energies, the spectrum is fitted in different $p_{T}$ ranges. The extracted values of $T_{\mathrm{eff}}$, shown in Fig.~\ref{Fig:dp} (right), seem to be consistent within collision energies for different fit ranges, which suggests common sources for direct photon production independent of the collision energy.

\section{Non-prompt direct photons}

\begin{figure}
\centerline{%
\includegraphics[height=5.25cm]{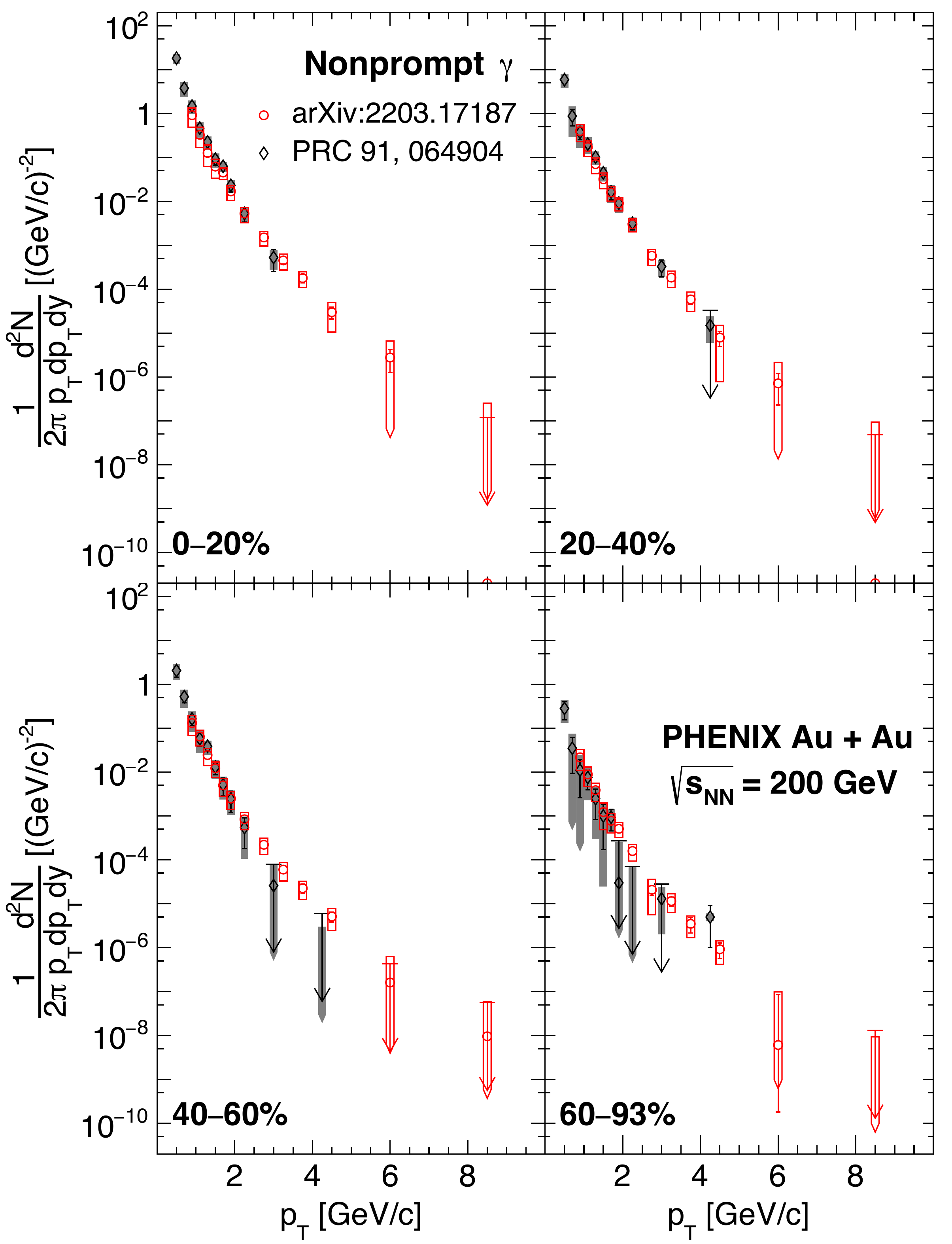}}
\caption{Invariant yield of non-prompt direct photons as a function of $p_{T}$ for Au+Au at \snn{200} for different collision centralities.}
\label{Fig:ndp_spec}
\end{figure}

Non-prompt direct photons are defined as the excess photons above the prompt photon contributions. These are radiations that are emitted during the collision from the hot and expanding fireball and are estimated by subtracting the $N_{\mathrm{coll}}$ scaled p+p fit from the direct photon spectrum. In Fig.~\ref{Fig:ndp_spec}, the non-prompt direct photon spectra are shown for every 20\% collision centrality. There has been a significant increase in the reach of the conversion photon measurement both in $p_{T}$ and in centrality as compared to previous publications.

\begin{figure}
\centerline{%
\includegraphics[height=3.8cm]{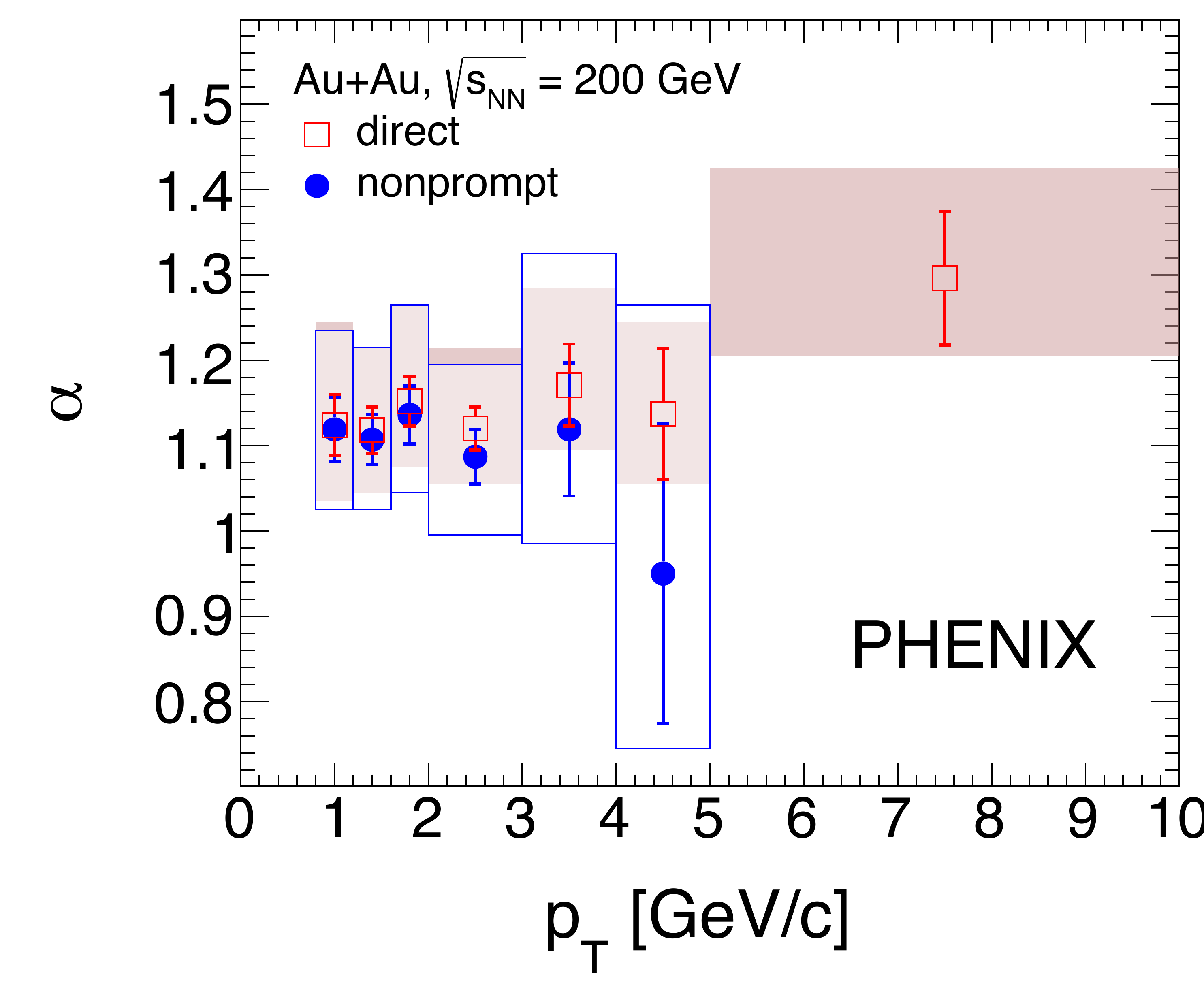}
\includegraphics[height=3.8cm]{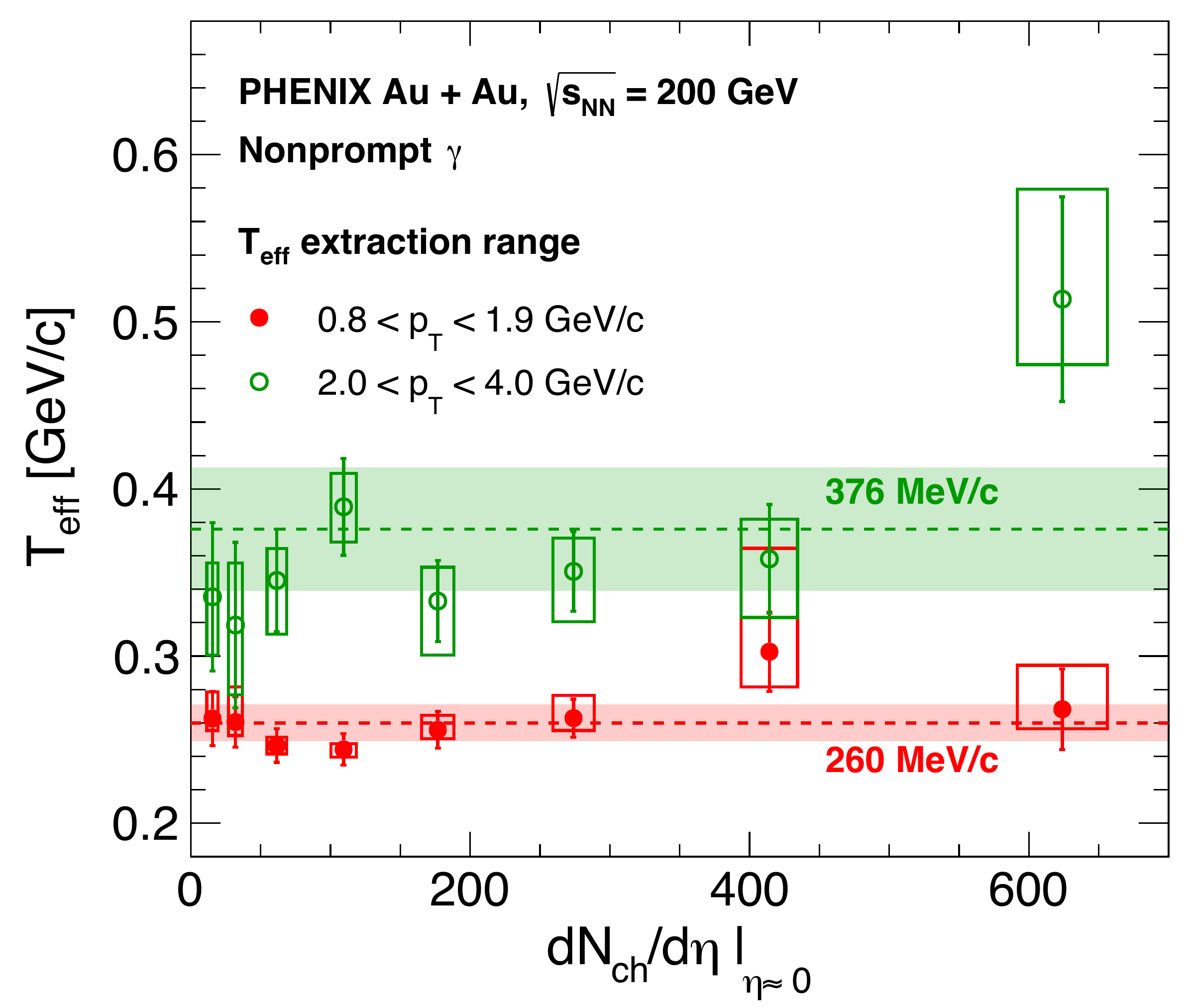}}
\caption{Scaling power $\alpha$, as a function of $p_{T}$ (left) and the inverse slope of non-prompt direct photon spectrum as a function of system size for different $p_{T}$ ranges (right).}
\label{Fig:npdp}
\end{figure}

\textbf{Universal scaling}: To study the centrality dependence, the scaling  power, $\alpha$, is extracted by fitting the $p_{T}$-integrated yield of non-prompt direct photon spectrum as a function of charged particle multiplicity at midrapidity for six non-overlapping $p_{T}$ ranges. In Fig.~\ref{Fig:npdp} (left), $\alpha$ as a function of $p_{T}$ is shown. 
It is compared with the $\alpha$ extracted from the $p_{T}$-integrated direct photon spectra in the same $p_{T}$ ranges. Below 3 GeV/$c$, the direct photon spectra are dominated by non-prompt direct photon sources and hence the $\alpha$ powers for both direct photons and non-prompt direct photons are similar. However, with increasing $p_{T}$, they start to differ, although it should be noted that the non-prompt direct photon spectra run out of statistics. Theoretical calculations suggest that $\alpha$ increases as we go from the Hadron Gas to the QGP phase~\cite{Shen:2013vja}, however, in experiments, we find $\alpha$ to be relatively independent of $p_{T}$.

\textbf{Effective temperature}: The shape of the non-prompt direct photon spectrum is not described by a single exponential but rather has a continuously increasing inverse slope with $p_{T}$. To quantify this changing slope, the non-prompt direct photon spectra is fitted with exponential in two different $p_{T}$ ranges, as shown in Fig.~\ref{Fig:npdp} (right), and the slopes are found to be consistent with a constant value and independent of the collision centrality. The average value of the slope increases from 200 MeV to about 400 MeV in the $p_{T}$ range of 0.8 to 4 GeV/$c$. A $T_{\mathrm{eff}}$ greater than 350 MeV suggests that we may be sensitive to sources beyond the Hadron Gas phase. The change $T_{\mathrm{eff}}$ is not surprising as the underlying spectra is time-integrated over the full evolution of the expanding fireball, from its earliest pre-equilibrium state through the QGP phase, crossing over to the hadron gas phase and further expanding and cooling till the hadrons eventually stop interacting among themselves. In turn, contributions from the earliest phase are likely to dominate the spectra at higher $p_{T}$, which is consistent with the observation of an increasing $T_{\mathrm{eff}}$ with $p_{T}$.

\section{Theoretical comparisons}

\begin{figure}
\centerline{%
\includegraphics[height=4.8cm]{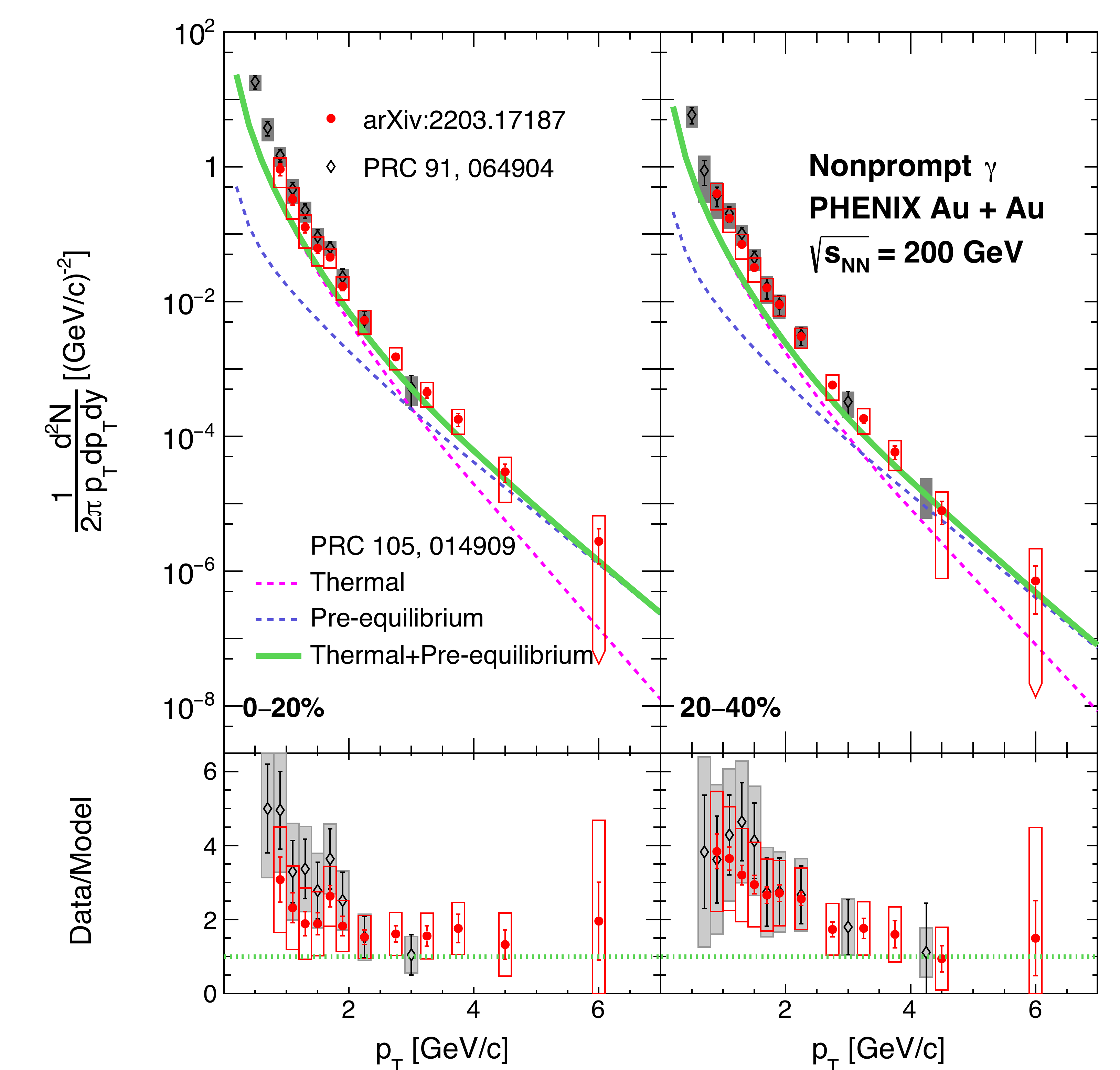}
\includegraphics[height=3.8cm]{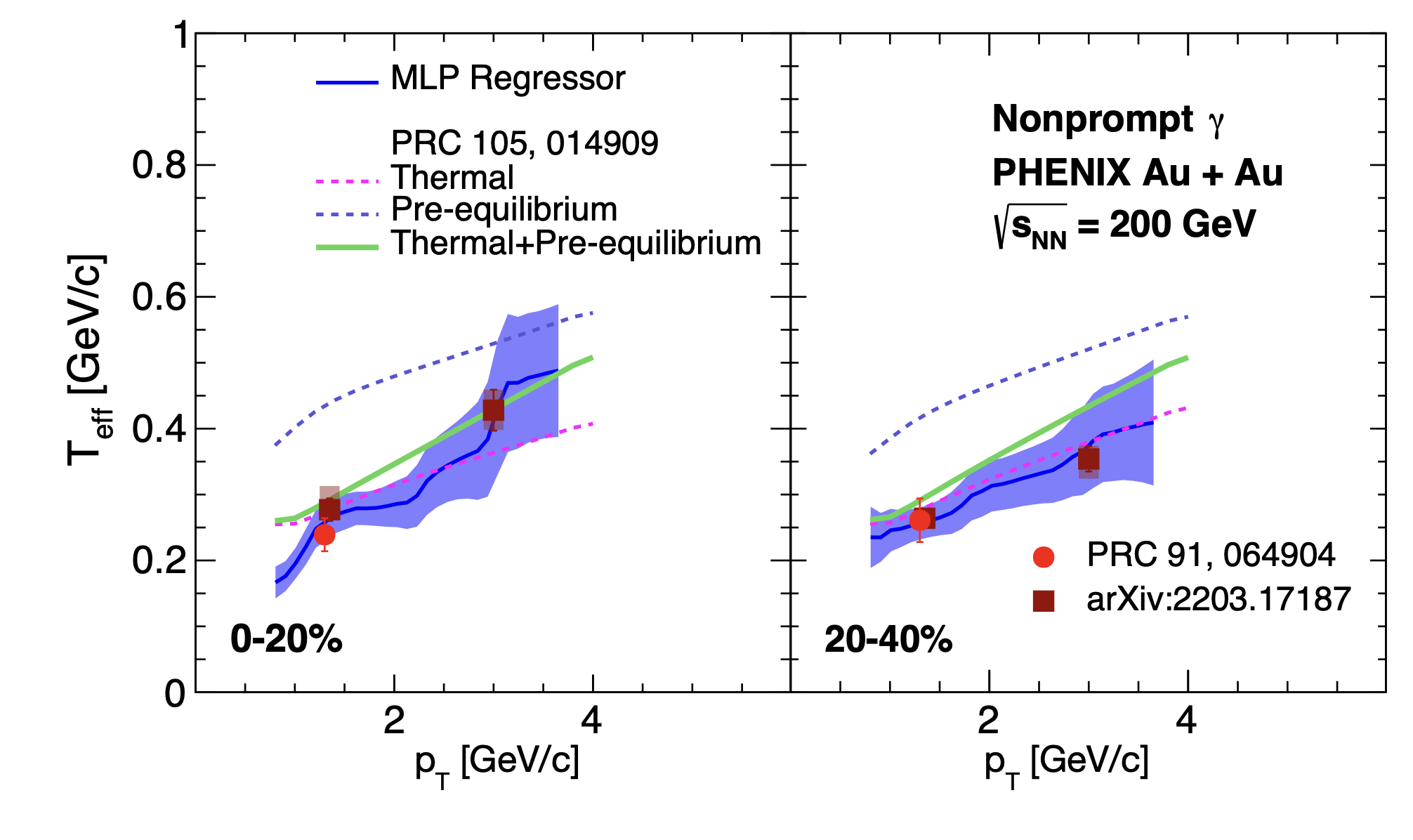}}
\caption{Comparison of the non-prompt direct photon spectrum for Au+Au at \snn{200} with theoretical calculations.}
\label{Fig:theory}
\end{figure}

The measured non-prompt direct photon spectra are compared to recent theoretical calculations which employ a hybrid model and the contributions from the pre-equilibrium state is highlighted on both photonic and hardonic observables in Fig.~\ref{Fig:theory} (left)~\cite{Gale:2021emg}. The bottom panel shows the ratio of the measurements with the combined thermal and pre-equilibrium contributions from theory. The calculations predict that the pre-equilibrium radiations become the dominant source above a $p_{T}$ of 3 GeV/$c$. While the shape is reproduced well, the overall yield predicted by the model falls short especially below 2 GeV/$c$ where the measured yields seem to be a factor 2--3 larger.

In order to further explore the shape of the non-prompt direct photon spectra, they are smoothened using a machine learning based regression algorithm called Multi Layer Perceptron on the PHENIX data~\cite{PHENIX:2022rsx}~\cite{PHENIX:2014nkk}. The inverse slope is extracted by numerical differentiation and is shown in Fig.~\ref{Fig:theory} (right). It can be argued that with increasing $p_{T}$, the contribution from the pre-equilibrium phase may be important.

\section{Summary}
To summarize, we discussed the results of the measurement of direct photons for Au+Au collisions at 39, 62.4 and 200 GeV. In addition, a more differential measurement with non-prompt direct photons is shown for the high statistics Au+Au collisions at 200 GeV. Universal scaling, independent of centrality, collision energy and collision system is observed with charged particle multiplicity at midrapidity. The scaling power, $\alpha$, is independent of $p_{T}$ for both direct and non-prompt photons. Both direct photons and non-prompt direct photon spectra exhibit an increasing inverse slope with $p_{T}$. Recent theoretical calculations including pre-equilibrium contributions seem to reduce the discrepancy between theory and observation. 

\bibliography{references}

\end{document}